#\documentclass{ptephy_v1}%%%%where ptephy_v1 is the template name
\documentclass[preprint]{ptephy_v1}%%%%%% to generate preprint number with ptep logo

\usepackage{hyperref}
\def\be{\begin{equation}}
\def\ee{\end{equation}}
\def\bea{\begin{eqnarray}}
\def\eea{\end{eqnarray}}
\def\bse{\begin{subequations}}
\def\ese{\end{subequations}}

\usepackage[T1]{fontenc}

\begin{document}

\title{Ground state correlations on ground state densities and total binding energies of  $^{40} \mathrm{Ca}$,  $^{48} \mathrm{Ca}$ and $^{208}$Pb%Insert the article title here
}

%%%% To generate auto affiliation numbers please use \author{}\affil{} command

\author{F. Minato}
\affil{Nuclear Data Center, Japan Atomic Energy Agency, Tokai, Ibaraki 319-1195, Japan
\email{minato.futoshi@jaea.go.jp}}
\affil{RIKEN Nishina Center, Wako, Saitama 351-0198, Japan}

\author{H.~Sagawa}%Insert second author name here}
\affil{RIKEN Nishina Center, Wako, Saitama 351-0198, Japan}
\affil{Center for Mathematical and Physics, the University of Aizu, Aizu-Wakamatsu, Fukushima 965-8580, Japan}
\author{S.~Yoshida}%Insert third author name here}
\affil{Science Research Center, Hosei University,
2-17-1 Fujimi, Chiyoda, Tokyo 102-8160, Japan}%Insert fourth author name here} %%% Use optional bracket [3] to change the respective address
%\affil{Insert third author address here}

%\author{Insert last author name here\thanks{These authors contributed equally to this work}}
%\affil{Insert last author address here}

%%% To include the collaborator name... Please use the command "\collaborator"
%%% For example: \collaborator{ATLAS Collaboration}

\begin{abstract}
Neutron and proton densities of doubly-closed shell nuclei  $^{40} \mathrm{Ca}$,  $^{48} \mathrm{Ca}$ and $^{208}$Pb are studied  based on a Hartree-Fock model with SAMi, SAMi-J27 and SAMi-T energy density functionals (EDFs). 
The ground state correlations (GSC) induced by isoscalar and isovector phonons are also evaluated by the second order perturbation theory  with a  self-consistent random phase approximation (RPA). We found that the interior part of the ground state densities is reduced by the  GSC in consistent with the experimental data.  
On the other hand, the GSC enhances the neutron skin thickness of  $^{48} \mathrm{Ca}$ and $^{208}$Pb. The effect of GSC on the total binding energy is also evaluated by the quasi-boson approximation. The effect of the tensor interaction is found small on both the density distributions and the binding energies.  
\end{abstract}
\subjectindex{ground state density, binding energy, Hartree-Fock approximation, ground state correlation}
\maketitle
\section{Introduction}%Insert A head here}
Mean field models have been quite successful to describe gross features of ground state mean square radii and the total binding energies of many nuclei in a wide region of the mass table~\cite{Goriely2016, MinShi2019, Kortelainen2014}. Recent studies of the density distributions of doubly-closed shell nuclei show appreciable differences between experimental and theoretical predictions not only in the interior part but also the surface region~\cite{Zenihiro,Zenihiro2}. 
In the Hartree-Fock (HF) model for the doubly-closed shell nuclei, the single-particle orbits are fully occupied up to the Fermi energy from the bottom of the potential and no occupation probabilities of the orbits above the Fermi energy.
On the other hand, it has been reported by experiments~\cite{p-trans1, p-trans2} and theoretical large-scale shell model calculations in $^{40}$Ca~\cite{Sagawa-PLB,Brown,Streets,Shimizu,Suzuki} that partial occupation probabilities of the orbits take values between $0$ and $1$, especially around the Fermi energy.
These partial occupations are mostly induced by the many-body correlations beyond the mean field models. 
\par
In this paper, we study the ground state correlations (GSCs) induced by phonons calculated by a self-consistent Hartree-Fock (HF)+random phase approximation (RPA) with Skyrme energy density functionals (EDFs).
The GSC are evaluated in the second-order perturbation model for the ground state densities, and the quasi-boson approximation for the total binding energies of $^{40} \mathrm{Ca}$, $^{48} \mathrm{Ca}$ and $^{208}$Pb.
A similar study has been performed in Ref.~\cite{Faessler1976, Waroquier1983} for $^{16}$O, $^{40,48}$Ca, and $^{208}$Pb.
This work, however, studies the GSC with a modern version of Skyrme parameters, SAMi~\cite{Roca-Maza2012}, SAMi-J27~\cite{Roca-Maza2013}, and SAMi-T~\cite{Shen2019} EDFs and also focuses on the effect of the tensor force.
\par
The paper is organized as follows. Section II is devoted to describe the theoretical model for GSC.  Numerical results are given in Section III.
Summary and future perspectives are given in Section IV. 
\section{Model for GSC}
The ground state density distributions of doubly-closed shell nuclei are commonly calculated by the Skyrme Hartree-Fock (SHF) model %with SAMi, SAMi-27, SAMi-T forces 
 assuming spherical symmetry.
The density distribution of radial component is given as
\begin{equation}
\rho_{q}(r)=\sum_{k=m,i} n_{k} \frac{2j_{k}+1}{4\pi} \left|\varphi_{k}(r)\right|^{2} \quad (q=p, \, n),
\label{eq:density}
\end{equation}
where $n_{k}$, $\varphi_{k}$ and $j_{k}$ are the occupation probability, the single-particle wave function of radial part and the total angular momentum for state $k$, respectively.
Here, we denote  $m$ for particle states and $i$ for hole states.
The occupation probabilities are $n_{m}=0$ and $n_{i}=1$ for the HF  density distribution.
In case of the perturbed density distribution, the  occupation probabilities are modified within the number-operator method~\cite{Rowe1968c} to be
\begin{equation}
\begin{split}
n_{m}&=\frac{1}{2(2j_{m}+1)} \sum_{i\lambda J\pi} (2J+1)\left|Y_{mi}(\lambda J^{\pi})\right|^2\\
n_{i}&=1-\frac{1}{2(2j_{i}+1)} \sum_{m\lambda J\pi} (2J+1)\left|Y_{mi}(\lambda J^{\pi})\right|^2.
\end{split}
\label{eq:occ}
\end{equation}
where $Y_{mi}(\lambda J^{\pi})$ is the backward amplitude of the phonon operator of random phase approximation. 
The label $\lambda$ and $J^{\pi}$ represent the  $\lambda$-th excited RPA phonon state and its spin-parity, respectively.
The RPA phonon creation operator is defined as
\begin{equation}
Q^{\dagger}(\lambda J^{\pi})=\sum_{mi} X_{mi}(\lambda J^{\pi})a_{m}^{\dagger}a_{i}-Y_{mi}(\lambda J^{\pi})a_{i}^{\dagger}a_{m}.
\label{eq:rpaph}
\end{equation}
The coefficients of $X_{mi}(\lambda J^{\pi})$ and $Y_{mi}(\lambda J^{\pi})$ are determined by solving the RPA equation~\cite{RingSchuck}.
\par
The rms radii are calculated by
\begin{equation}
    r_{q}\equiv \sqrt{\langle r^2 \rangle_{q}} =\sqrt{\frac{1}{N_{q}}4\pi \int \rho_{q}(r) r^{4}\,dr},
\end{equation}
where $N_{p}=Z$ and $N_{n}=N$. The neutron skin thickness is estimated by $\Delta r_{np}=r_{n}-r_{p}$.
\par
To calculate the unperturbed and perturbed density distributions numerically, we adopt the SHF model in the coordinate space.
To solve the RPA equation, we take into account  the single particle states up to $60$~MeV in energy, and restrict unperturbed $1p$-$1h$ excitations energy up to $100$~MeV 
in the model.
The continuum states are discretized by a finite box with a size of $R=20$~fm. 
We consider the spin-parity states up to $J^{\pi}=5^{\pm}$.  We confirm that the model space is enough to make  the calculated occupation probabilities of Eq.~\eqref{eq:occ} converge. 
The residual interactions, the second-derivative of SHF EDF with respect to densities, are self-consistently taken into account in the RPA except for spin-orbit component, which is however not significant.
Within this model space, the energy-weighted sum rules are satisfied over $99.5$\% for all $J^{\pi}$.
%
%
%
%
%
%%% 6pm July 28th, 2022. %%
\section{Results}
\subsection{Density distributions and rms radii}
%\section{Nuclear matter and ground$-$state properties calculated by using  RMF Lagrangians  and Skyrme parameters}
\begin{table*}[tb]
\begin{center}
\caption{Proton and neutron rms radii of $^{40} \mathrm{Ca}$, $^{48} \mathrm{Ca}$, and $^{208}$Pb.
The neutron skin is given as $\Delta r_{np}\equiv r_n-r_p$. 
HF denotes the results of HF calculations, while HF+GSC includes also the ground state correlations.
The experimental data are taken from~\cite{Zenihiro} for $^{40} \mathrm{Ca}$, $^{48} \mathrm{Ca}$ and from ~\cite{Zenihiro2} for $^{208}$Pb.
The  experimental errors of $\Delta r_{np}$ include both statistical errors and model uncertainties.
The radii are given in unit of fm.}
\label{tab1}
  \begin{tabular}{lc|cc|cc|cc|c}%D{.}{.}{1}D{.}{.}{2}D{.}{.}{2}D{.}{.}{2}D{.}{.}{2}D{.}{.}{2}|D{.}{.}{3}D{.}{.}{3}D{.}{.}{3}}
%    & & & & & & & \multicolumn{3}{c}{${}^{40} \mathrm{Ca}$} %& \multicolumn{3}{c} {$^{208}$Pb}
%    \\ 
    \hline
  & &\multicolumn{2}{c}{SAMi} & \multicolumn{2}{c}{SAMi-J27} & \multicolumn{2}{c}{SAMi-T} & %\multicolumn{1}{c}{$K_{\mathrm{sym}}$ (MeV)} & \multicolumn{1}{c}{$K_{\tau}$ (MeV)} & \multicolumn{1}{c|}{$m_{\mathrm{eff}}/m$} & \multicolumn{1}{c}{$r_p$ (fm)} & \multicolumn{1}{c}{$r_n$ (fm)} & \multicolumn{1}{c}{$\Delta r_{np}$ (fm)} %& $r_p$ (fm) & $r_n$ (fm) & $\Delta r_{np}$ (fm)
    \\ \hline
  Nuclei       &   & HF & HF+GSC &HF & HF+GSC &HF & HF+GSC & exp.  \\ \hline
 &  $r_n$ &  3.343 & 3.773 & 3.344 &3.810 & 3.339 & 3.808 & 3.375 \\
 $^{40}$Ca & $r_p$ & 3.390 &  3.853 & 3.391 & 3.870 & 3.386 & 3.845 & 3.385 \\
   &$\Delta r_{np}$  & $-$0.047 & $-$0.080&  $-$0.047 & $-$0.060 & $-$0.046 &$-$0.037 & $-0.010^{+0.049}_{-0.048}$ \\\hline
    & $r_n$ & 3.612 & 4.003 &  3.588 & 4.006 & 3.589 & 4.004 & 3.555 \\
   $^{48}$Ca &  $r_p$&  3.436 & 3.770 & 3.444 & 3.801 & 3.424 & 3.781 & 3.387 \\
    & $\Delta r_{np}$  & 0.176 & 0.233 & 0.144 & 0.205 & 0.166 & 0.223 & $0.168^{+0.052}_{-0.055}$ \\\hline
     & $r_n$ & 5.610 & 5.750& 5.580 & 5.731 & 5.574 &5.730 & 5.653 \\
    $^{208}$Pb &  $r_p$&   5.463 & 5.555   & 5.456 &5.553 & 5.421 & 5.517 & 5.442 \\
      &  $\Delta r_{np}$  & 0.147 & 0.195 & 0.123 & 0.178 & 0.153 & 0.213 & $0.211^{+0.054}_{-0.063}$ \\\hline
   \end{tabular}
   \end{center}
\end{table*}
Table~\ref{tab1} shows nuclear proton and neutron rms radii of $^{40} \mathrm{Ca}$, $^{48} \mathrm{Ca}$ and $^{208}$Pb calculated by using   a modern version of Skyrme parameters, SAMi, SAMi$-$J27, and SAMi-T EDFs.
Both the results of HF and HF+GSC calculations are listed and compared with experimental data.  
%Figure  \ref{fig1} shows the calculated neutron and proton densities of $^{40}$Ca by using 
%SAMi EDFs.   
The tensor interactions are included in SAMi-T EDF.    %The experimental proton and neutron densities are also shown in Figure  \ref{fig1}. 
The proton radii or equivalently the charge radii of $^{40} \mathrm{Ca}$ and $^{208}$Pb are in general included in the optimized data set. Therefore it is not a surprise that HF results reproduce well the proton radii of these nuclei. The neutron radii of $^{40} \mathrm{Ca}$ and $^{208}$Pb are slightly smaller than the experimental ones.  
Including the GSC, the neutron skin of 
$^{208}$Pb increases and becomes closer to the experimental value compared with the HF result.  On the other hand, the neutron skin of $^{40} \mathrm{Ca}$ becomes a larger negative value with the GSC in the cases of SAMi and SAMi-J27, while SAMi-T with the tensor term gives slightly better agreement with the experimental value $\Delta r_{np}=-0.01$~fm.  
In $^{48} \mathrm{Ca}$,  all the three HF results in Table~\ref{tab1} give slightly larger  radii than the experimental one for both protons and neutrons, while the neutron skin is well reproduced by the HF 
calculations, especially in the case of SAMi-T.  The GSC enlarges both the proton and the neutron radii and the quantitative agreement becomes worse than  HF results by about 0.4 fm.
%For example, in $^{48} \mathrm{Ca}$, the neutron skin is well accounted by HF, and the GSC overestimates the skin by $0.04$--$0.07$~fm.
Experimental data of the neutron skin of $^{208}$Pb is still controversial. The dipole polarizability experiment gives $\Delta r_{np}=0.156^{+0.025}_{-0.021}$~fm~\cite{Tamii}, while the parity violation electron scattering experiment PREXII were analyzed by two different models and the neutron skin was extracted to be $\Delta r_{np}=0.283\pm0.071$~fm~\cite{PREXII} and $\Delta r_{np}=0.19\pm0.02$~fm 
\cite{PG-PREX}, respectively.  The present calculations, both HF and HF+GSC, are unlikely to reproduce the experimental data of PREXII  reported  in ref. \cite{PREXII}. 

\begin{figure}[tb]
\begin{center}
% \includegraphics[clip,width=1.0\linewidth,bb=0 250 550 750]{%Pb$-$isgmr$-$Ex.pdf}%
%fig1s.pdf}
 \vspace*{3.0cm}
 \includegraphics[width=0.9\linewidth,bb=0  200 600 700]%bb=0 0 482 312
{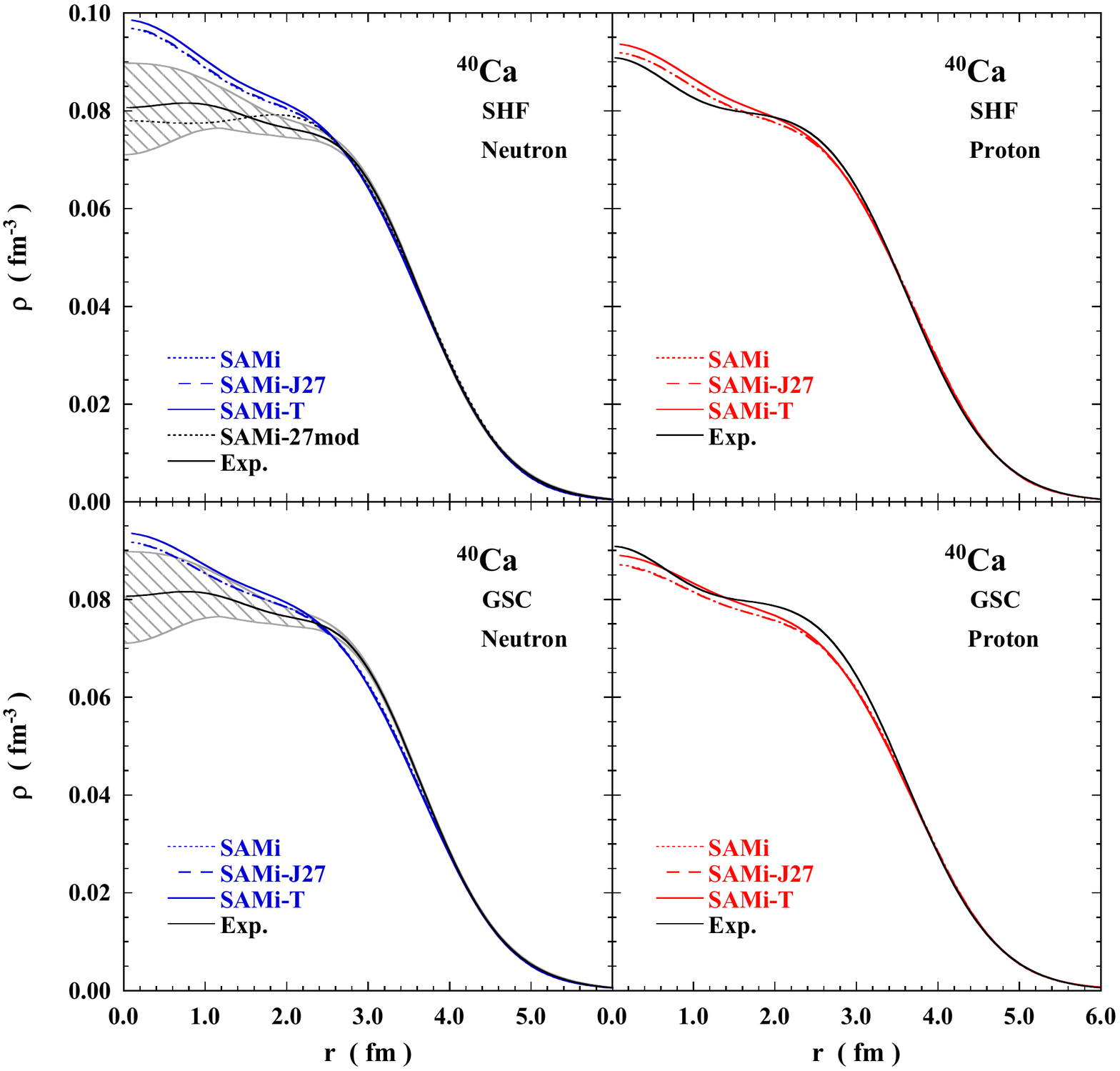}
%\includegraphics[width=0.8\linewidth,bb=0  0 795 452]%bb=0 0 482 312
%ca40-SHF-den.pdf}
%\includegraphics[width=0.8\linewidth,bb=0  0 795 452]%bb=0 0 482 312
%{ca40(GSC)-den-new.pdf}
  \vspace*{-1.0cm}
\caption{(color online) 
Neutron (left panels) and  proton (right panels) density distributions  of $^{40} \mathrm{Ca}$.
Upper panels show the HF results with three SAMi EDFs, SAMi, SAMi-J27 and SAMi-T, respectively.  
The SAMi-T has the tensor terms, while the SAMi and SAMi-J27 have no tensor terms.  
% together with calculated ones using SAMi  interactions. 
%For a guide to eyes, the neutron density is shifted by 0.02 fm$^{-3}$.  
%Neutron (proton) densities are plotted by solid (dashed) lines.  
%The left panel shows  the neutron density,  while the right panel is the proton density. 
The black solid lines show experimental data  taken from Ref.~\cite{electron-sca} for protons and from Ref.~\cite{Zenihiro} for neutrons. 
The shaded area of experimental neutron density shows experimental uncertainties of both statistical and systematic errors. The neutron density distribution denoted SAMi-J27mod in the upper left panel is obtained by optimizing the occupation probabilities of neutrons as given in Table~\ref{tab2}.
 \label{fig1}}
\end{center}
\end{figure}
\begin{figure}[ht]
\begin{center}
%\includegraphics[clip,width=1.0\linewidth,bb=0 250 550 750]{%Pb$-$isgmr$-$Ex.pdf}%
%fig1s.pdf}
\vspace*{-2.0cm}
\includegraphics[width=1.\linewidth,bb=0  00 600 600]{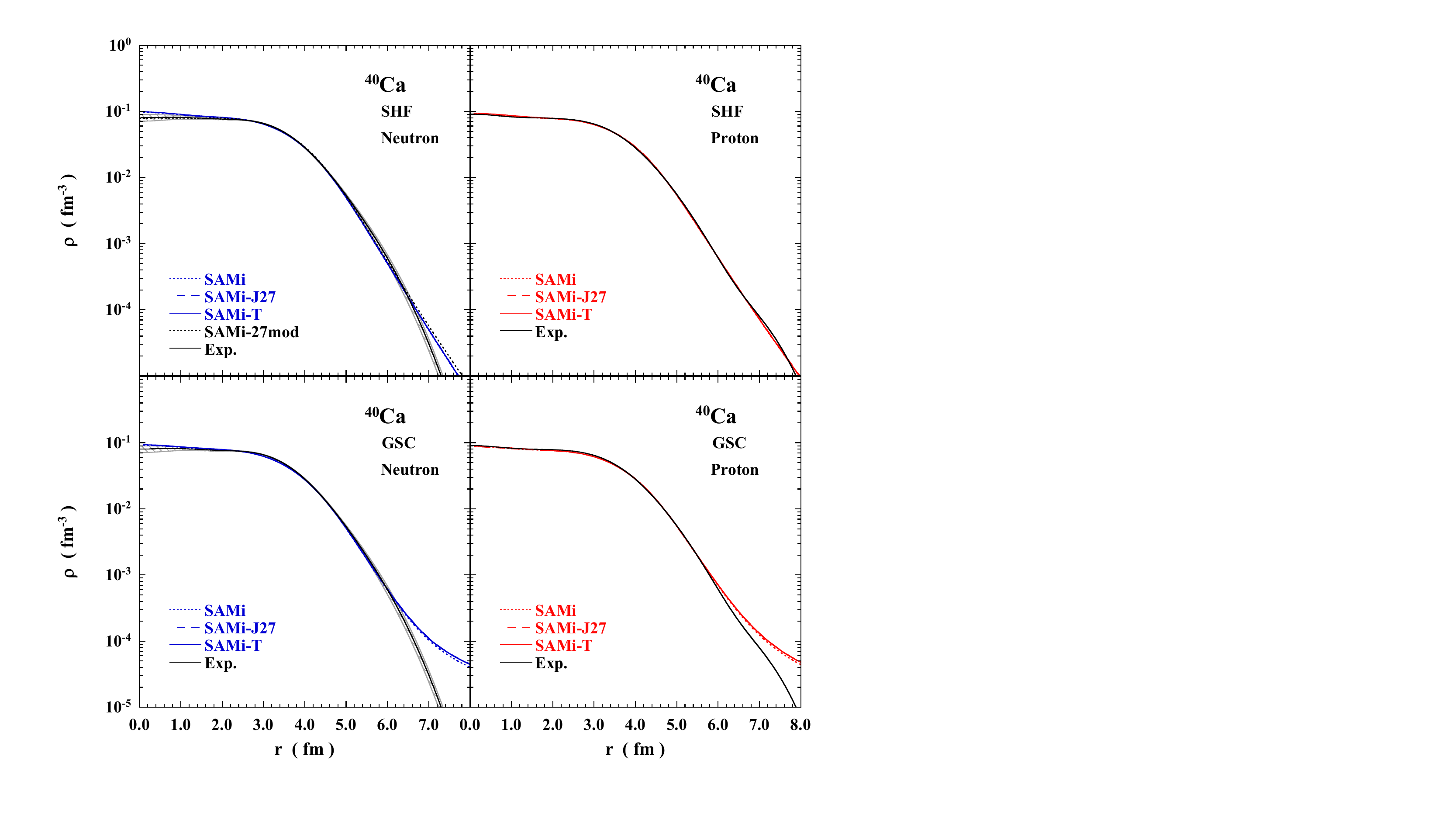}
\vspace*{-1.0cm}
\caption{(color online) 
Neutron and proton density distributions of $^{40} \mathrm{Ca}$ in the log scale.
The upper panels show the HF results, while the lower panels are the HF+GCS results. Experimental data are shown by a black solid curve. The model dependence of EDFs is not visible in this scale.}
\label{ca40-den-log}
\end{center}
\end{figure}
Figure~\ref{fig1} shows comparisons between calculated and experimental neutron (left panel) and proton (right panel) densities of $^{40}$Ca.  Upper panels show the HF results, and the lower panels correspond to the results of HF+GSC.  The HF densities of both neutrons and protons  somewhat overestimate in the interior part.  The EDF model dependence is rather small: the SAMi and SAMi-J27 give almost identical results for the density distributions, while SAMi-T enhances the interior region of density more than other two  interactions.  On the contrarily, the calculated results  slightly underestimate the shoulder part at around $3$~fm.   With the GSC, the central part is hindered and become closer to the experimental density distribution, especially in the case of proton density.  On top of that, the shoulder part is also quenched and small amount of density is shifted to $r>6$~fm region. In the upper left panel of Fig. \ref{fig1},  the neutron density distribution denoted SAMi-J27mod is shown by optimizing the occupation probabilities of neutrons as given in Table~\ref{tab2} in order to reproduce the experimental density. %  in Fig. 1. 
 %The optimized density is    denoted as SAMi-27mod.  
The difference between HF and HF+GSC at $r>6.5$~fm is clearly seen in Fig. \ref{ca40-den-log} where the densities are plotted in log scale.  In this scale,
the interaction dependence does not appear in the entire region of coordinate space.
\par
\begin{figure}[tb]
\begin{center}
%\includegraphics[clip,width=1.0\linewidth,bb=0 250 550 750]{%Pb$-$isgmr$-$Ex.pdf}%
%fig1s.pdf}
%\vspace*{2.0cm}
\includegraphics[width=0.9\linewidth,bb=0  0 800 500]{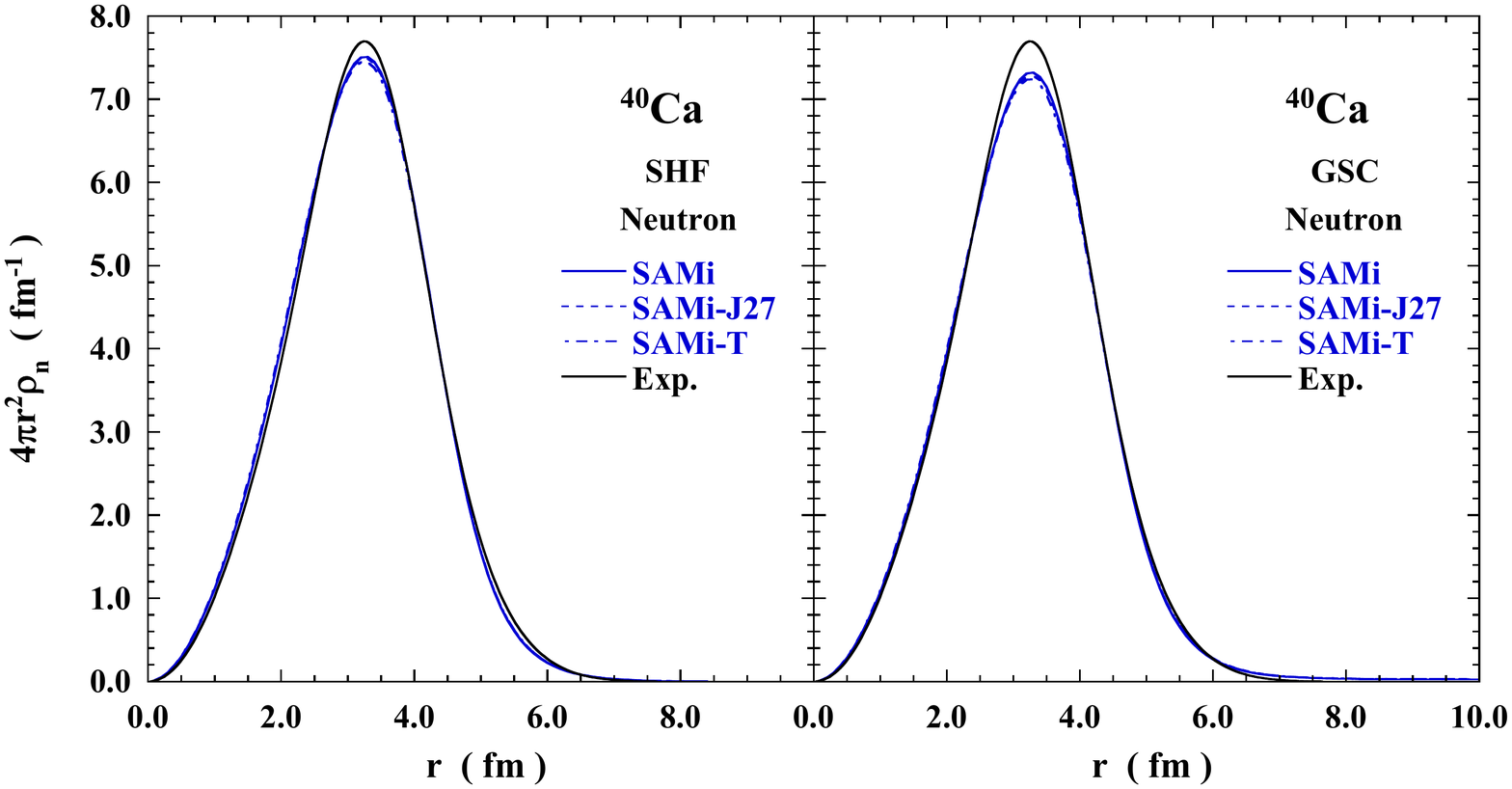}%ca40-n-r2-density.pdf}
\vspace*{-1.5cm}
\caption{(color online) 
Neutron density distributions of $^{40} \mathrm{Ca}$ with a factor of $4\pi r^2$.  
The left panel shows the HF results, while the right panel are the HF+GCS results. Experimental data are shown by a black solid curve. The model dependence of EDFs is not visible in this scale.}
\label{fig1a}
\end{center}
\end{figure}

Figure ~\ref{fig1a} shows the neutron density distribution of $^{40}$Ca without and with GSC multiplied by a factor $4\pi r^{2}$. 
%Neutron density distributions of $^{40}$Ca are shown in Fig.~\ref{fig1a} without and with GSC multiplying a factor $4\pi r^{2}$ on the density $\rho(r)$.  
The differences between HF and HF+GSC appear in the central region $r<2$~fm, the peak height at $r=3.2$ fm, in the tail region 5 fm$<r<$6 fm, and also at the very low density region $r>6.5$~fm. 
We can see small improvements of agreements by GSC at the central region and also  at the surface.
On the other hand, the peak height is quenched by GSC and a small amount of density is pushed to outside of the surface region in contrast to the experimental data.

\begin{figure}[tb]
\begin{center}
%\vspace*{-1.0cm}
\vspace*{3.0cm}
 \includegraphics[width=0.9\linewidth,bb=0  200 600 700]%bb=0 0 482 312
{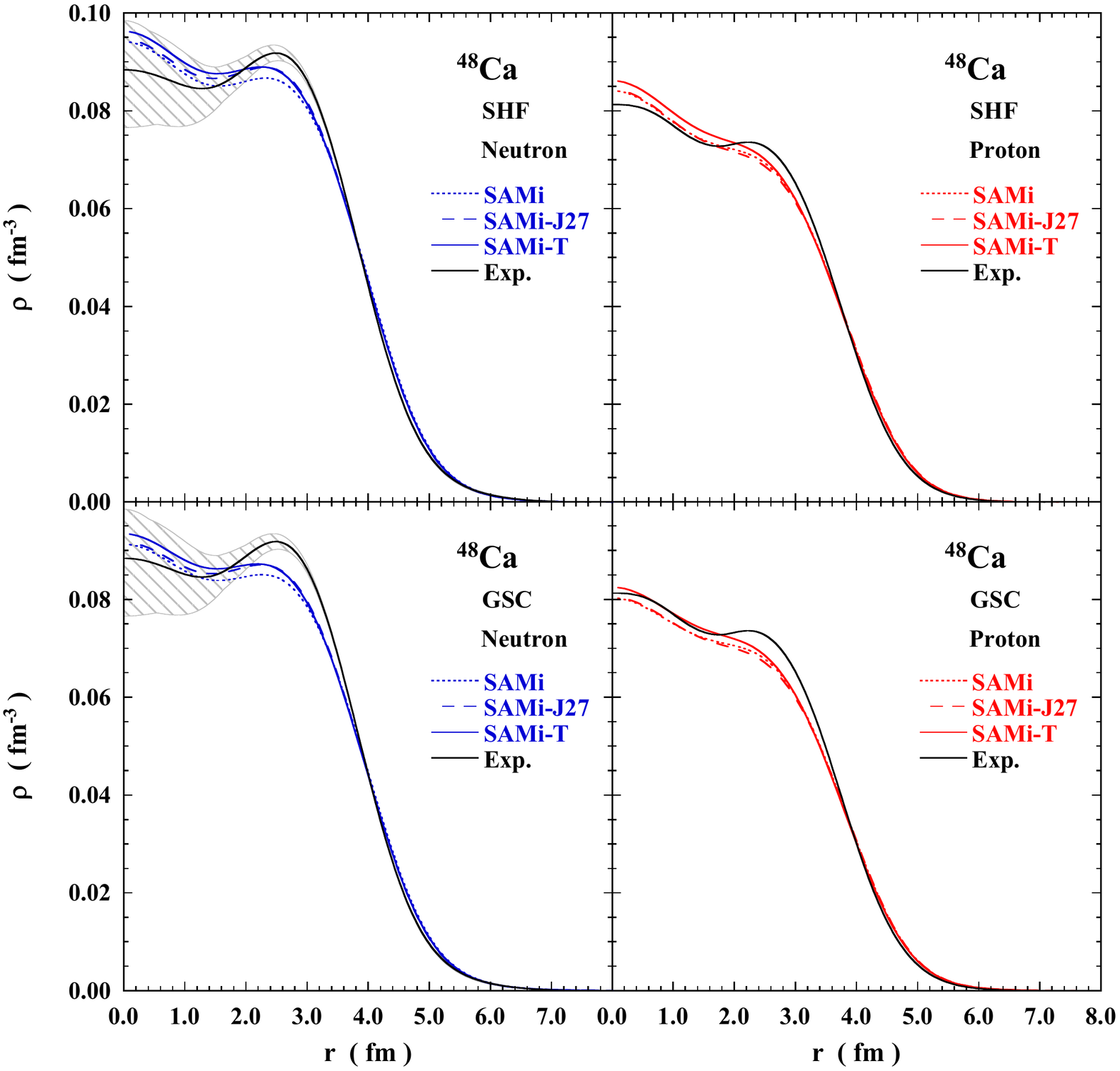}
\vspace*{-1.0cm}
\caption{(color online) 
The same as Fig.~\ref{fig1}, but for $^{48} \mathrm{Ca}$.}
\label{fig2}
\end{center}
\end{figure}
%\vspace*{-1.0cm}
%
Figure~\ref{fig2} shows comparisons between calculated and experimental densities of neutrons (left panel) and protons (right panel)  of $^{48}$Ca. 
The essential features are the same as those of $^{40}$Ca.
% Upper panels show HF results, and the lower panels correspond to results of HF+GSC.  
One peculiar difference is the shoulder peak of neutron density due to the $f_{7/2}$ neutron occupation in $^{48}$Ca.
The HF densities of both neutrons and protons   slightly overestimate in the interior part.  The EDF model dependence is again rather small for $^{48}$Ca. %: the SAMi and SAMi-27 give almost identical results for the density distributions, while SAMi-T enhance the interior region on density more than other two  interactions.  
The calculated results   underestimate the shoulder part at around $3$~fm. 
With the GSC, the central part is hindered and become closer to the experimental density distribution, or even underestimate in the case of proton density.  The shoulder part is also quenched and small amount of density is shifted to $r>6$~fm region.
\begin{figure}[tb]
\begin{center}
\vspace*{3.0cm}
 \includegraphics[width=0.9\linewidth,bb=0  200 600 700]%bb=0 0 482 312
{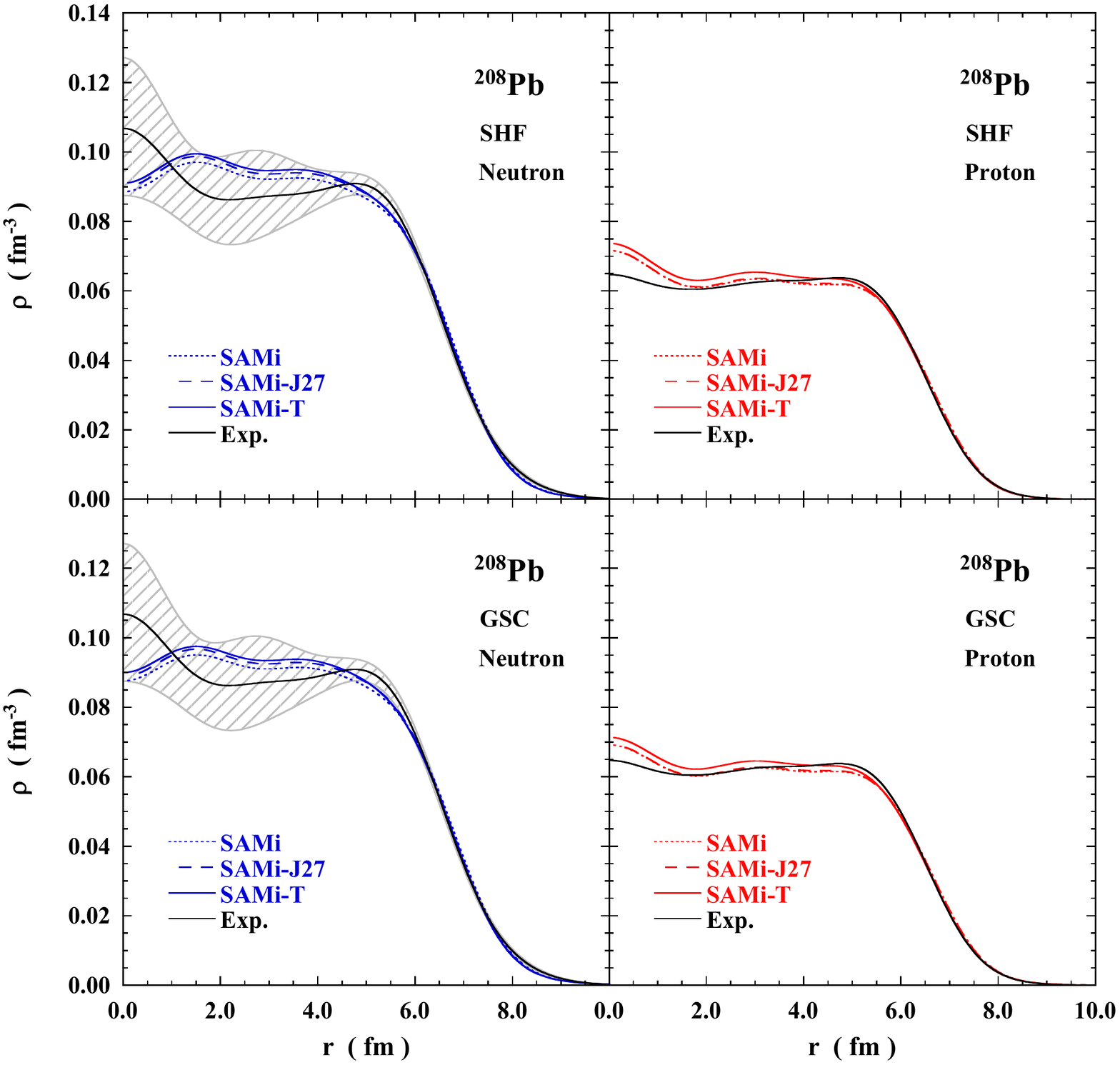}
  \vspace*{-1.0cm}
\caption{(color online) 
The same as Fig.~\ref{fig1}, but for $^{208} \mathrm{Pb}$. 
Experimental data  taken from Ref.~\cite{electron-sca} for protons and from Ref.~\cite{Zenihiro2} for neutrons.} \label{fig3}
\end{center}
\end{figure}
The densities of $^{208}$Pb are shown in Fig.~\ref{fig3}.   For the proton density (right panel),  the HF results reproduce well
the plateau between $r=2.0$~fm and $r=5.5$~fm, while they show larger values at the central part at $r<1.5$~fm than the experimental data. 
The GSC effect on the proton density of $^{208}$Pb is relatively small even compared with $^{40}$Ca and $^{48}$Ca.
Nevertheless, we can see that the HF+GSC improves slightly the agreement at the interior part of proton density at the right panels of Fig.~\ref{fig3}. 
For the neutron density, the HF results give slightly larger density for the plateau between $r=1.0$ and $r=5.5$~fm, while underestimate the central part $r<1$~fm. 
These disagreements are somewhat improved in the case of HF+GSC results and the calculated results are within the experimental uncertainties at $r<5$~fm.  he density distribution at $r > 8$ fm becomes larger than that of HF similar to the case of $^{40}$Ca shown in  Fig. \ref{ca40-den-log}.
%There are not appreciable changes by the GSC in the density distribution at $r>8$~fm.
%There are 
 %   a small variation %s in calculated densities 
 %   among difference different DDME$-$J effective Lagrangians. 
%
%
%
%
%
\subsection{Particle occupation numbers of $sd$-shell  and $pf$-shell orbits in $^{40}$Ca}
In Table~\ref{tab2}, the occupation probabilities of various calculations are listed with the 
empirical proton and neutron occupation probabilities in $^{40}$Ca.  
Skyrme HF-Bogolyubov calculations  are performed with SAMi interaction and the volume-type and mixed-type pairing interactions with the code HFBTHO~\cite{Sagawa-PLB}.
We found that there are essentially no occupation probabilities $v_j^2$ in $pf$-shell orbits because of a large shell gap between $1d_{3/2}$ and 
$1f_{7/2}$ orbits.    
The GSCs with SAMi and SAMi-T give small amounts of the occupation probabilities in the $pf$-shell orbits, 
$\sum_{pf}(2j+1)v_j^2 \approx$ 0.3. There is no essential difference between SAMi and SAMi-T without and with the tensor terms. 

The results of several Large Scale Shell Model (LSSM) calculations are also listed in Table~\ref{tab2}, which depend largely on the model space and  
 the adopted interactions. 
%Shell model results depend largely on the model space and the adopted interactions.    
In Refs.~\cite{Brown,Shimizu,Suzuki}, LSSM calculations have been performed including many-particle many-hole excitations from $sd$-shell to $pf$-shell configurations.
% The particle occupation numbers are summarized in Table \ref{tab2}.  
In Ref.~\cite{Brown}, the active model space is $(1d_{3/2}, 1f_{7/2}, 2p_{3/2})$, while 
the $1d_{3/2}, 2s_{1/2}$ orbits are also included in Ref.~\cite{Shimizu}.  Shell model calculations with the configurations of  2-particle 2-hole (2p-2h) and 4p-4h excitations from the closed shell core of $^{40}$Ca are also performed in Ref.~\cite{Suzuki}.   
The full $sd$ and $pf$ shell orbits  are involved with $sdpf$-mu shell model effective interactions.  The occupation probabilities of $1f$-orbits go up to $\sum_{f}(2j+1)v_j^2\approx 0.6$, but those of $p$-shell orbits are rather small as less than 0.1.   
  %The results of HF+GSC with SAMi and SAMi-T are
 % also listed in Table \ref{tab2}.
The empirical occupation numbers are also listed obtained from proton transfer reactions~\cite{p-trans1}, and also from the analysis of neutron density distributions  denoted as
SAMi-J27mod in Fig.~\ref{fig1}~\cite{Sagawa-PLB}. The proton transfer data suggest the same amount of the proton occupation probabilities in $pf$-shell orbits as the shell model predictions. On the other hand, the experimental neutron density distribution requires the occupation probability of
$f$-orbits to be $\sim$1.2, which is two times larger than the proton transfer experiments.
\begin{table*} [t]
  \caption{Particle occupation numbers, $(2j+1)v_j^2$, for neutron $sd$-shell and $pf$-shell configurations of $^{40}$Ca. The column with a bar (---) is not involved in the shell model calculations or  analysis of experimental data.
  Experimental data of proton transfer reaction are taken from Ref.~\cite{p-trans1}.}
\label{tab2}
\begin{center}
  \begin{tabular}{l|cccccc|cccc}%l|D{.}{.}{1}D{.}{.}{2}D{.}{.}{2}|D{.}{.}{2}D{.}{.}{2}D{.}{.}{2}|D{.}{.}{3}%D{.}{.}{3}}%D{.}{.}{3}}
%    & & & & & & & \multicolumn{3}{c}{${}^{40} \mathrm{Ca}$} %& \multicolumn{3}{c} {$^{208}$Pb}
%    \\ 
    \hline
   Model & $1s_{1/2}$ & $2p_{3/2}$ &$2p_{1/2}$ &$1d_{5/2}$ & $2s_{1/2}$ & $1d_{3/2}$ &  $1f_{7/2}$ &  $2p_{3/2}$ &$2p_{1/2}$ & $1f_{5/2}$   \\\hline 
   % \multicolumn{1}{c}{$1d_{5/2}$} & \multicolumn{1}{c}{$2s_{1/2}$ &\multicolumn{1}{c} $1d_{3/2}$ & \multicolumn{1}{c}{$1f_{7/2}$} & \multicolumn{1}{c|}{c & \multicolumn{1}{c}{$2p_{1/2}$} & \multicolumn{1}{c}{$1f_{5/2}$} \\\line %& \multicolumn{1}{c}{$\Delta r_{np}$ (fm)} 
HFB (SAMi)  & 2.00 & 4.00 & 2.00 &6.00  &2.00  &4.00  &   0.00 & 0.00 & 0.00 & 0.00  \\ 
HF+GSC(SAMi) & 1.964 & 3.908 &1.949&5.761 & 1.843 &3.792 & 0.206 & 0.049 & 0.010 &  0.040 \\
HF+GSC(SAMi-T) &1.968 & 3.905 & 1.954 &5.753 & 1.845 & 3.776 & 0.201 & 0.052 & 0.009 & 0.076 \\
GSC (RPA)~\cite{Lenske} & 1.92 & 3.76 & 1.88 & 5.58 & 1.8 & 3.7 &     0.24 & 0.16  & 0.08 &--- \\
 $dpf$-shell~\cite{Brown} & --- & --- &--- & --- & --- & 3.30 & 0.63 & 0.07 & ---  & --- \\
  $sdpf$-msd4~\cite{Shimizu} & --- & --- &--- & 5.902 & 1.908 & 3.477 & 0.617 & 0.096 & --- & --- \\
  $sdpf$-mu $(2p$-$2h)$~\cite{Suzuki} & --- & --- &--- &  5.864 & 1.933 & 3.845 & 0.191 & 0.040 & 0.020 & 0.107 \\
  $sdpf$-mu $(4p$-$4h)$~\cite{Suzuki}  & --- & --- &--- &  5.727& 1.854 & 3.660 & 0.421 & 0.089 & 0.041 & 0.208 \\
  exp. (p-transfer)~\cite{p-trans1} & --- & --- &--- & 6.0 & 1.70& 3.59 & 0.56  & 0.15 &  &  \\
  SAMi-J27mod~\cite{Sagawa-PLB}& 2.00 & 4.00 & 2.00 & 6.00   & 1.342  & 3.431 &  1.227 & &  &  \\\hline
  \end{tabular}
 \end{center}
\end{table*}   
\subsection{Correlation energies for the total binding energies}
The correlation energies due to the ground state correlations are calculated within the framework of RPA and quasi-boson approximation~\cite{Rowe1968c}:
\be  \label{eq-corr}
E_{\rm corr}=\sum_{\lambda J \pi} (2J+1) E(\lambda J^{\pi}) \sum_{mi} \left|Y_{mi}(\lambda J^{\pi})\right|^2.
\ee
The sum of phonon states $\lambda$ in Eq.~\eqref{eq-corr} is restricted to those absorb a fraction of the non-energy-weighted sum rule larger than $1$\% for isoscalar or isovector transition strength.
%The correlarton energies due to the ground state correlations are calculated by using the second-order perturbation theory formula:
%\be  \label{eq-corr}
%E_{\rm corr}=  ***.
%\ee
The results are tabulated in Table~\ref{tab-corr}. As is expected, the HF calculations give already reasonable agreement with the experimental total binding energies since all the parameters are optimized to reproduce a set of experimental data in which the total binding energies of the doubly-closed shell nuclei are included. % always with highest priority.  
The correlation energies give about $1$\% additional energies to the binding energies of $^{40}$Ca and $^{48}$Ca, while it is much smaller in $^{208}$Pb to be about $0.2$\%. 
As seen in the table, the correlation energies do not give further improvement. 
For this reason, together with the rms radii, one needs to readjust the Skyrme parameters including the correlation effects for the consistent description of the ground state properties in the beyond mean field model.
\begin{table*} [tb]
\begin{center}
  \caption{The HF total binding energies and correlation energies due the ground state correlations of $^{40,48} \mathrm{Ca}$\ and $^{208}$Pb. %\textcolor{red}{
  The experimental data are taken from AME2020~\cite{AME2020}.%}
  The correlation energies are calculated by using Eq. \eqref{eq-corr}. The value is given by a unit of MeV.}
\label{tab-corr}
\begin{tabular}{r|rr|rr|rr|r}%D{.}{.}{1}D{.}{.}{2}D{.}{.}{2}D{.}{.}{2}D{.}{.}{2}D{.}{.}{2}|D{.}{.}{3}D{.}{.}{3}D{.}{.}{3}}
%    & & & & & & & \multicolumn{3}{c}{${}^{40} \mathrm{Ca}$} %& \multicolumn{3}{c} {$^{208}$Pb}
%    \\ 
    \hline
  & \multicolumn{2}{c}{SAMi} & \multicolumn{2}{c}{SAMi-J27} & \multicolumn{2}{c}{SAMi-T} &   %\multicolumn{1}{c}{$K_{\mathrm{sym}}$ (MeV)} & \multicolumn{1}{c}{$K_{\tau}$ (MeV)} & \multicolumn{1}{c|}{$m_{\mathrm{eff}}/m$} & \multicolumn{1}{c}{$r_p$ (fm)} & \multicolumn{1}{c}{$r_n$ (fm)} & \multicolumn{1}{c}{$\Delta r_{np}$ (fm)} %& $r_p$ (fm) & $r_n$ (fm) & $\Delta r_{np}$ (fm)
    \\ \hline
  Nuclei       &    E$_{\rm tot}$ & E$_{\rm corr}$ &E$_{\rm tot}$ & E$_{\rm corr}$  & E$_{\rm tot}$ & E$_{\rm corr}$ & exp. \\ \hline
% & n &  3.343 & 3.773 & 3.344 &3.810 & 3.339 & 3.808 & 3.375 \\
 $^{40}$Ca &  347.02 &  4.25 & 344.39 & 4.10 & 343.68 & 4.78  & 342.05 \\
%   &$\Delta r_{np}$  & $-$0.047 & $-$0.080&  $-$0.047 & $-$0.060 & $-$0.046 &$-$0.037 & $-$0.010 \\\hline
%    & n& 3.612 & 4.003 &  3.588 & 4.006 & 3.589 & 4.004 & 3.555 \\
   $^{48}$Ca &   415.46 & 4.93 & 415.45  & 5.89 & 416.73 & 4.76    & 416.00\\
 %%   & $\Delta r_{np}$  & 0.176 & 0.233 & 0.144 & 0.205 & 0.166 & 0.223 & 0.168 \\\hline
%%     & n& 5.610 & 5.750& 5.580 & 5.731 & 5.574 &5.730 & 5.653 \\
    $^{208}$Pb & 1635.86 &   3.95 & 1635.54 & 3.05 & 1635.39 & 3.41  & 1636.43 \\
 %     &  $\Delta r_{np}$  & 0.147 & 0.195 & 0.123 & 0.178 & 0.153 & 0.213 & 0.211 
  \hline
     \end{tabular}
   \end{center}
\end{table*}
\section{Summary and future perspectives}
We studied the effects of GSC on the densities and the total binding energies of doubly-closed shell nuclei $^{40}$Ca, $^{48}$Ca, and $^{208}$Pb with modern Skyrme EDFs, SAMi, SAMi-J27 and SAMi-T. For the ground state densities, the GSC reduces slightly in the interior part of HF densities of both neutrons and protons, and gives consistent results in comparisons with the experimental data of $^{40}$Ca and $^{48}$Ca.
The GSC decreases also the shoulder part of density at around $r=3$~fm, and removes a small amount of density to the region $r>6$~fm.  
The GSC effect is smaller in $^{208}$Pb and quantitatively the same as $^{40}$Ca, and $^{48}$Ca.  
\par
The GSC on the total binding energy is also discussed within the quasi-boson approximation, and found to increase about $1$\% for $^{40,48}$Ca and $0.2$\% for $^{208}$Pb of the total binding energy. 
In order to describe these ground state observables with beyond mean field models, we need to refit Skyrme EDFs including the GSC effects for the consistent descriptions.  
\section*{Acknowledgement}
FM and HS thank YIPQS long-term workshop 
``Mean-field and Cluster Dynamics in Nuclear 
Systems 2022 (MCD2022)" May 9 -- June 17, 2022 held at Yukawa Institute, Kyoto University,  where the collaboration was started.
We thank J. Zenihiro and T. Uesaka for providing the experimental data. 
We thank also T. Naito for useful discussions on the HF calculations. This work is supported by JSPS KAKENHI Grant Number JP19K03858.JSPS.
\par


\begin{thebibliography}{99}
\bibitem{Goriely2016}
S.~Goriely, N.~Chamel, and J.M.~Pearson,
Phys. Rev. C 93, 034337 (2016).
%
\bibitem{MinShi2019}
Min~Shi, Zhong-Ming~Niu, and Hao-Zhao~Liang,
Chin. Phys. C 43, 074104 (2019).
%
\bibitem{Kortelainen2014}
M.~Kortelainen, J.~McDonnell, W.~Nazarewicz, E.~Olsen, P.-G.~Reinhard, J.~Sarich, N.~Schunck, S.M.~Wild, D.~Davesne, J.~Erler, and A.~Pastore,
Phys. Rev. C 89, 054314 (2014).
%
\bibitem{Zenihiro} 
J.~Zenihiro, T. Uesaka, S. Yoshida and H. Sagawa,  Prog, Theo. Exp. Phys. 2021, 023D05 (2021)\\
J.~Zenihiro et al., arXiv:1810.11796 (2018). 
%
\bibitem{Zenihiro2} 
J.~Zenihiro  {\it et al}., Phys. Rev. C82, 044611 (2010).
%
\bibitem{p-trans1} 
P. Doll, G.J.~Wagner, K.T.~Kn\"{o}pfle  and G. Mairle, 
Nucl. Phys. A 263, 210 (1976).
%
\bibitem{p-trans2}
F. Malaguti, A. Uguzzoni, E. Verondini  and P. E. Hodgson,
Nucl. Phys. A 297, 287 (1978); ibid., Nuovo Cim. A 49, 412 (1979).
%
\bibitem{Brown} 
B. A.~Brown, S. E.~Massen, and P. E.~Hodgson,  J. Phys. G: Nucl. Phys. 5,  1655  (1979).
%
\bibitem{Streets}
J.~Streets, B.A.~Brown and P.E.~Hodgson, J. Phys. G: Nucl. Phys. 8,   839 (1982).
%
\bibitem{Shimizu} 
N. Shimizu, Y. Utsuno, T. Ichikawa et al., private communications (2022).
%
\bibitem{Suzuki} 
Toshio Suzuki, private communications (2022).
%
\bibitem{Sagawa-PLB}
H. Sagawa, S. Yoshida, T. Naito, T. Uesaka, J. Zenihiro, J. Tanaka, Physics Letters B829 137072 (2022).
%
\bibitem{Faessler1976}
A.~Faessler, S.~Krewald, A.~Plastino, and J.~Speth,
Z. Physik A 276, 91 (1976).
%
\bibitem{Waroquier1983}
M.~Waroquier, J.~Bloch, G.~Wenes, and K.~Heyde,
Phys. Rev. C 28, 1791 (1983).
%
\bibitem{Roca-Maza2012}
X.~Roca-Maza, G.~Col\`{o}, and H.~Sagawa,
Phys. Rev. C 86, 031306(R) (2012).
%
\bibitem{Roca-Maza2013}
X.~Roca-Maza, M.~Brenna, B. K.~Agrawal, P. F.~Bortignon, G.~Col\`{o}, Li-Gang~Cao, N.~Paar, and D.~Vretenar,
Phys. Rev. C 87, 034301 (2013).
%
\bibitem{Shen2019}
S.~Shen, G.~Col\'{o}, and X.~Roca-Maza,
Phys. Rev. C 99, 034322 (2019).
%
\bibitem{Rowe1968c} 
D.J.~Rowe, Phys. Rev. 175, 1283 (1968).
%
\bibitem{RingSchuck} 
P.~Ring and P.~Schuck, 
{\it The Nuclear Many-Body Problem}, Springer-Verlag,  New York, 1980.
%
\bibitem{Lenske} 
H.~Lenske and J.~Wambach,  Phys. Lett. B 249, 377 (1990).
%
\bibitem{Tamii} 
A.~Tamii et al., Phys. Rev. Lett. 107, 062502 (2011).
%
\bibitem{PREXII} 
D.~Adhikari et al. (PREX Collaboration), Phys. Rev. Lett. 126, 172502 (2021).
%
\bibitem{PG-PREX}
P.-G.~Reinhard, X.~Roca-Maza, and W.~Nazarewicz
Phys. Rev. Lett. 127, 232501 (2021). 
%
\bibitem{electron-sca}
H. de Vries, C. W. de Jager, and C. de Vries,
At.~Data Nucl.~Data Tables~\textbf{36}, 495 (1987).
%
\bibitem{AME2020}
W. J.~Huang, M.~Wang, F.G.~Kondev, G.~Audi, and S.~Naimi,
Chin. Phys. C45, 030002 (2021);
Chin. Phys. C45, 030003 (2021).
%
\end{thebibliography}
\end{document}